\documentclass[sigconf]{acmart}

\usepackage{booktabs} 

\setcopyright{rightsretained}
\usepackage{amsmath,amssymb,amsfonts}
\usepackage{algorithmic}
\usepackage{graphicx}
\usepackage{textcomp}
\usepackage{booktabs}
\usepackage{subcaption}
\usepackage{siunitx}
\usepackage[ruled]{algorithm2e} 
\usepackage{enumitem}

\usepackage{url}

\acmDOI{10.475/123_4}

\acmISBN{123-4567-24-567/08/06}

\acmConference[BuildSys]{ACM conference}{2018}{Shenzhen, China}
\renewcommand{\baselinestretch}{1}

\acmYear{1997}
\copyrightyear{2016}

\acmArticle{4}
\acmPrice{15.00}

\begin{document}
\title{Demo Abstract: Pible: Battery-Free Mote for Perpetual Indoor BLE Applications}

\author{Francesco Fraternali}
\orcid{1234-5678-9012-3456}
\affiliation{%
  \institution{University of California, San Diego}
  }
  \email{frfrater@ucsd.edu}
\author{Bharathan Balaji}
\affiliation{%
  \institution{University of California, Los Angeles}
}
\email{bbalaji@ucla.edu }

\author{Yuvraj Agarwal}
\affiliation{%
  \institution{Carnegie Mellon University}
  }
\email{yuvraj@cs.cmu.edu }  
  
\author{Luca Benini}
\affiliation{%
  \institution{University of Bologna - ETH Zurich}
  }
\email{luca.benini@unibo.it }  
  
\author{Rajesh Gupta}
\affiliation{%
  \institution{University of California, San Diego}
}
\email{gupta@eng.ucsd.edu }

\renewcommand{\shortauthors}{F. Fraternali et al.}

\begin{abstract}
As of today, large-scale wireless sensor networks are adopted for smart building applications as they are easy and flexible to deploy. Low-power wireless nodes can achieve multi-year lifetimes with an AA battery using Bluetooth Low Energy (BLE) and ZigBee. However, replacing these batteries at scale is a non-trivial, labor-intensive task. Energy harvesting has emerged as a potential solution to avoid battery replacement but requires compromises such as application specific sensor node design, simplified communication protocol or reduced quality of service. We show the design of a battery-free sensor node using commercial off the shelf components, and present Pible: a Perpetual Indoor BLE sensor node that uses an ambient light energy harvesting system and can support numerous smart building applications.
We show trade-offs between node-lifetime, quality of service and light availability and present a predictive algorithm that adapts to changing lighting conditions to maximize node lifetime and application quality of service. 
\end{abstract}

%
%

\ccsdesc[500]{Computer systems organization~Sensor networks}
\ccsdesc[500]{Computer systems organization~Embedded systems}

\keywords{Smart Buildings, Wireless Sensor Network, Energy-Harvesting}

\acmYear{2018}\copyrightyear{2018}
\setcopyright{acmcopyright}
\acmConference[BuildSys '18]{BuildSys '18: Conference on Systems for Built Environments}{November 7--8, 2018}{Shenzen, China}
\acmBooktitle{BuildSys '18: Conference on Systems for Built Environments, November 7--8, 2018, Shenzen, China}
\acmPrice{15.00}
\acmDOI{10.1145/3276774.3276785}
\acmISBN{978-1-4503-5951-1/18/11}

\maketitle

\section{Introduction}

Buildings are integrated with thousands of sensors and the sensing systems are designed for wired communication and power during the design of building itself. 
The wired infrastructure comes at the cost of rigidity and changes can be prohibitively expensive, e.g., retrofitting of a single wired thermostat can cost \$2500 \cite{link:retrofitting-cost}. 

Wireless sensors have emerged as the answer to this cost/rigidity problem. With low power and low data rates communication protocols such as ZigBee, 6LowPAN and Bluetooth Low Energy (BLE), wireless sensors can be deployed with a multi-year battery lifetime. 
But these sensor nodes are powered by batteries that require periodic manual replacement.

As we scale to large deployments, the manual replacement of batteries becomes a bottleneck. 
Battery replacements can be mitigated using energy harvesting: DoubleDip measures water flow powering itself using temperature difference~\cite{paper:DoubleDip}, Monjolo harvests energy from power lines and measures plug-level power consumption~\cite{paper:Monjolo}, 
and ambient light has been used for indoor monitoring applications \cite{paper:solar_feasibility}. 
Despite these innovations, there are limited commercial devices that use indoor energy harvesting solutions. We highlight 3 limitations that inhibit adoption: (i) they are designed for specific applications, (ii) they do not support standard protocols like ZigBee or BLE, (iii) the application quality of service (QoS) is inadequate. 

With advances in low power microcontrollers, integrated radios, and systems on chip, we show the feasibility of overcoming these limitations using commercial off the shelf components. We explore the design space of a generic energy harvesting sensor node for indoor monitoring applications with the objective of perpetual operations in typical environments.
We designed and built \emph{Pible}, a Perpetual Indoor BLE sensor node that can be used for a wide set of building applications that span from periodic sensor measurements, i.e. temperature, light, to event-driven sensors, i.e. PIR, door events and BLE beaconing. We show trade-offs between QoS, lifetime and harvested energy that enables our prototype sensor node to work in different lighting conditions inside buildings. We introduce hardware solutions to increase charging efficiency and overcome cold-start operations that limit system functionality and usability. Finally, we propose a local sensor-node power management solution that adapts and maximizes the application-QoS and node-lifetime. 

\vspace{-2mm}


\section{Pible-Design}
\label{sec:Architecture}
We want to support common building applications such as (i) sensing environmental indoor conditions i.e. temperature, light, (ii) event-driven sensing including door sensors and motion sensors such as passive infra-red (PIR) sensors, and (iii) BLE beaconing. 
\vspace{-2mm}

\subsection{Hardware}
We use a general energy harvesting architecture for Pible \cite{Pible,Scaling_my}, shows Pible architecture: an energy harvester gathers and transfers power to an energy management board, that collects the gathered power to a storage element. Once the energy accumulated reaches a usable voltage level, the energy management board powers the microcontroller (MCU) that starts its operations. 

    

\subsubsection{Platform System on Chip, Antenna and Sensors}
We select the TI CC2650 chip that supports multiple communication protocols (e.g. 6LoWPAN, BLE). It consumes 1 \si\micro A in standby mode and can be connected to low-power MEMS sensors. We equip our board with temperature, light, humidity, pressure, reed switch, accelerometer, gyroscope, and a PIR motion sensor. 

\vspace{-2mm}
\subsubsection{Energy Storage}
\label{battery}
To increase lifetime, we adopt a super-capacitor as it supports up to 1 billion recharges~\cite{link:supercap_2}. We select a super-capacitor from Panasonic~\cite{link:supercap}, with capacitance 1F at 5.5V.

\vspace{-2mm}
\subsubsection{Energy Harvester}

We use solar light since it has high power density inside buildings. We use the indoor solar panel AM-1454 from Sanyo since it harvests 46.5 \si\micro A at 1.5V with 300 lux. 

\vspace{-2mm}
\subsubsection{Energy Management Board (EMB)}

We select the BQ25570 
that includes 2 programmable DC/DC converters: (i) an ultra-low-power boost converter (V$_{BAT}$) that is highly efficient when the storage element voltage level is above 1.8V and (ii) a nano-power buck converter (V$_{buck}$) that supports output current up to 110mA. 
We switch between the chargers to avoid `cold-start' operations. 

\vspace{-2mm}
\subsubsection{Wireless Communication Protocol}
We use Bluetooth Low Energy (BLE) as it offers several advantages for indoor environments~\cite{paper:Cosero}. It has a range of about 30 meters inside buildings.



\begin{figure}
  \centering
     \includegraphics[width=0.8\linewidth]{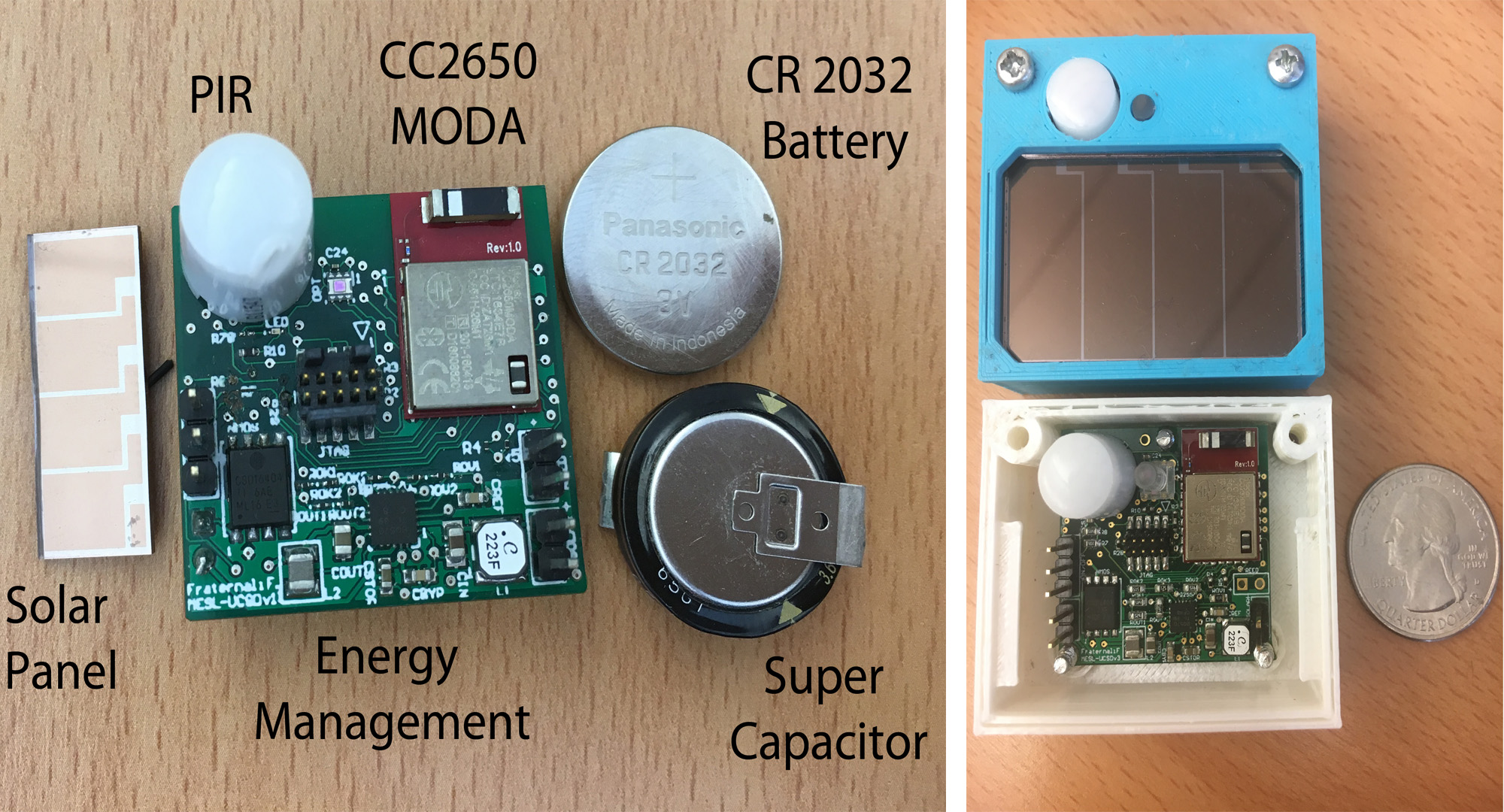}
     \vspace{-4mm}
    \caption{Pible: Perpetual Indoor BLE Node} 
		\vspace{-4mm}
    \label{fig:Pible}
\end{figure}

\vspace{-3mm}
\subsection{Software}

To facilitate scalability and ease deployment, we design our algorithm to be computational independent of external devices like the base station. Our algorithm uses a simple sensor specific lookup table and a lighting availability prediction to set the sensing rate. All the algorithmic decisions are made by the Pible MCU, Algorithm \ref{alg:one} shows the pseudo code.

\begin{algorithm}[ht]
\caption{Power Management Algorithm}
\label{alg:one}

\SetAlgoNoLine
\begin{algorithmic}[1]
\small
\STATE \textbf{Input:} Light-vec(5); Volt-vec(5); light=0; volt=0; index=0
\WHILE{true}
		\STATE volt = read SC Volt, light = read Light
        \STATE \textbf{if}(index==0) \text{or} (volt=max) \textbf{then} Next-QoS=based on voltage and Table~\ref{tab:QoS};  index++
        \STATE update Light[] with light; update Volt[] with volt
        \STATE \textbf{if}(light==0) \text{or} (Light trend$<$0) \textbf{then} --Next-QoS \textbf{else} ++Next-QoS
        
        \STATE \textbf{if} (Volt trend$=<$0) \text{and} (volt$\neq$max)
        	 \textbf{then} --Next-QoS \textbf{else} ++Next-QoS
   
        \STATE update QoS; sleep until next wakeup event\;
      
\ENDWHILE
\end{algorithmic}
\end{algorithm}

\begin{table}
\small
\centering
  \caption{Voltage Level and QoS Relationship}
  \label{tab:QoS}
  \vspace{-2mm}
  \begin{tabular}{cccccc}
    \toprule
    QoS  & Voltage & QoS & QoS-PIR & QoS\\
    State & [V] & Sensings [s]& Detection [s] & Advertising [s] \\
    \midrule
    7 & 3.6 - 3.4& 20 & 10 & 0.1\\
    6 & 3.4 - 3.2& 40 & 20 &0.2\\
    5 & 3.2 - 3.0& 60 & 30 &0.4\\
    4 & 3.0 - 2.8& 120 & 60& 0.64\\
    3 & 2.8 - 2.6& 240 &120& 0.9\\
    2 & 2.6 - 2.4& 300 &300& 2\\
    1 & 2.4 - 2.1& 600 &600& 5\\
  \bottomrule
  \vspace{-1em}
\end{tabular}
\end{table}
\vspace{-2mm}




\section{Demo Description}
\label{sec:Conclusion}
\begin{figure}
  \centering
     \includegraphics[width=0.7\linewidth]{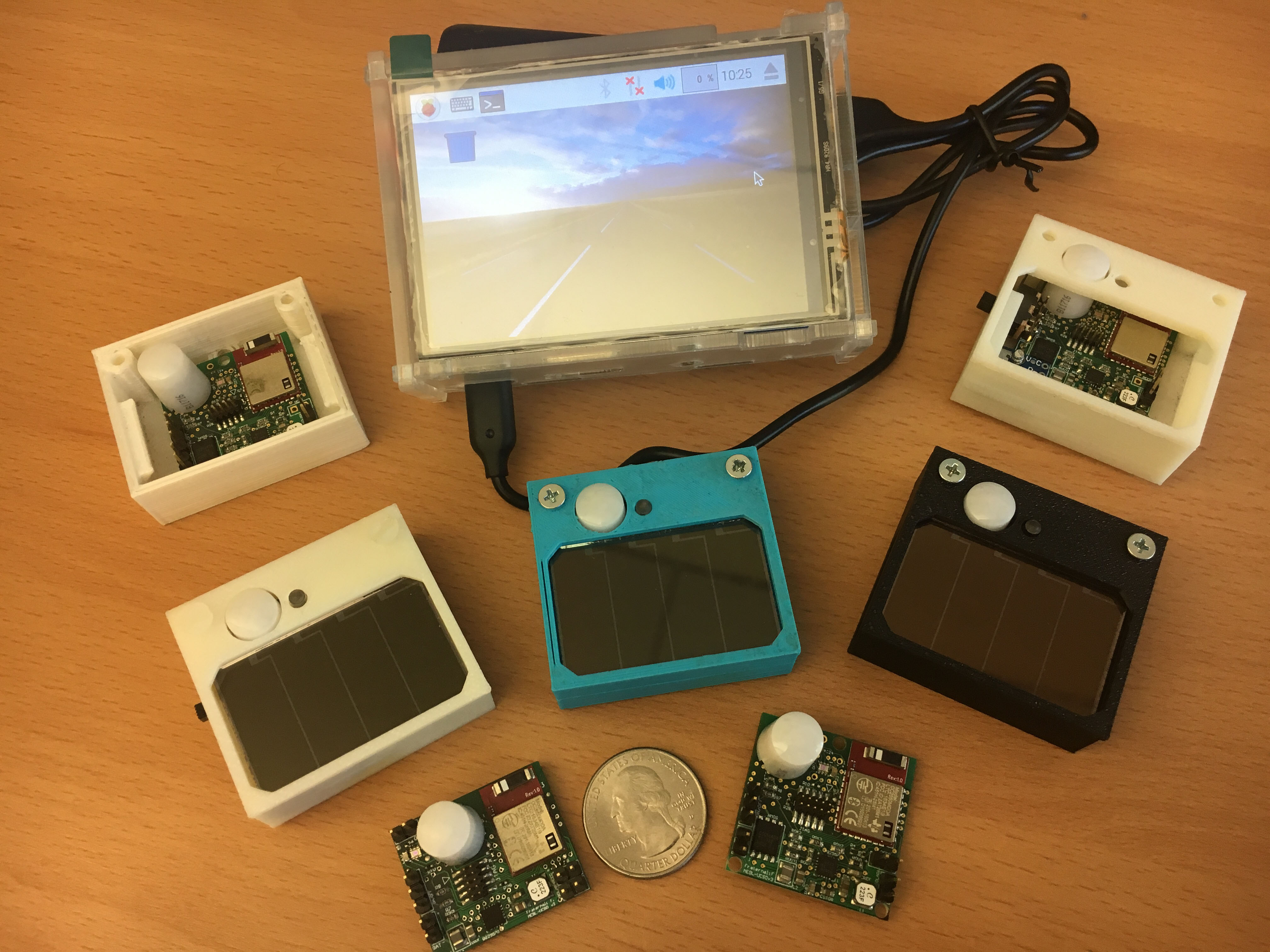}
      \vspace{-4mm}
    \caption{Main Components of our Demo}
    \label{fig:demo}
    \vspace{-2mm}
\end{figure}

All Pible-nodes deployed send data packet to the closest Base Station (BS). In our system, the Base Station is a Raspberry PI equipped with a BLE USB dongle. To monitor the nodes's status, each Pible node send to the BS information such as sensor data, QoS state and voltage level.
A picture of the system is reported in Figure \ref{fig:demo}.

During the demo, we will display in real-time the data packets received from each node and show how the power management algorithm adapts the sensing rate to increase lifetime while maximizing the quality of service applications. As future work, we will explore Reinforcement Learning to increase light patterns prediction.
\vspace{-2mm}

\section*{Acknowledgments}

This work is supported by the National Science Foundation grants CSR-1526237, TWC-1564009 and BD Spokes 1636879

\bibliographystyle{ACM-Reference-Format}
\bibliography{BuildSys-2018}

\end{document}